\documentclass[10pt,a4paper,english,conference]{IEEEtran}
\usepackage[T1]{fontenc}
\usepackage{verbatim}
\usepackage{textcomp}
\usepackage{amsmath}
\usepackage{amsthm}
\usepackage{amssymb}
\usepackage{graphicx}

\makeatletter

\pdfpageheight\paperheight
\pdfpagewidth\paperwidth

\theoremstyle{plain}
\newtheorem{thm}{\protect\theoremname}
\theoremstyle{plain}
\newtheorem{lem}[thm]{\protect\lemmaname}

\usepackage{subfigure}
\usepackage{epstopdf}
\usepackage{balance}
\usepackage{cite}
\usepackage{citesort}
\IEEEoverridecommandlockouts

\makeatother

\usepackage{babel}
\providecommand{\lemmaname}{Lemma}
\providecommand{\theoremname}{Theorem}

\begin{document}

\title{What Is the True Value of Dynamic TDD: \\
A MAC Layer Perspective
}

\author{{\normalsize{}Ming Ding$^{\ddagger}$, David L$\acute{\textrm{o}}$pez
P$\acute{\textrm{e}}$rez$^{\dagger}$}\emph{\normalsize{}, }{\normalsize{}Guoqiang
Mao$^{\nparallel}{}^{\ddagger}$}\emph{\normalsize{}, }{\normalsize{}Zihuai
Lin}\textit{\normalsize{}$^{\mathparagraph}$}\emph{\normalsize{}}\\
\textit{\small{}$^{\ddagger}$Data61, Australia }{\small{}\{Ming.Ding@data61.csiro.au\}}\textit{\small{}}\\
\textit{\small{}$^{\dagger}$Nokia Bell Labs, Ireland }{\small{}\{david.lopez-perez@nokia.com\}}\textit{\small{}}\\
\textit{\small{}$^{\nparallel}$School of Computing and Communication,
University of Technology Sydney, Australia}\\
\textit{\small{}$^{\mathparagraph}$The University of Sydney, Australia}
}
\maketitle
\begin{abstract}
Small cell networks (SCNs) are envisioned to embrace dynamic time
division duplexing (TDD) in order to tailor downlink (DL)/uplink (UL)
subframe resources to quick variations and burstiness of DL/UL traffic.
The study of dynamic TDD is particularly important because it provides
valuable insights on the full duplex transmission technology, which
has been identified as one of the candidate technologies for the 5th-generation
(5G) networks. Up to now, the existing works on dynamic TDD have shown
that the UL of dynamic TDD suffers from severe performance degradation
due to the strong DL-to-UL interference in the physical (PHY) layer.
This conclusion raises a fundamental question: Despite such obvious
technology disadvantage\emph{, what is the true value of dynamic TDD?
}In this paper, we answer this question from a media access control
(MAC) layer viewpoint and present analytical results on the DL/UL
time resource utilization (TRU) of \emph{synchronous} dynamic TDD,
which has been widely adopted in the existing 4th-generation (4G)
systems. Our analytical results shed new light on the dynamic TDD
in future synchronous 5G networks.%
\footnote{To appear in IEEE GLOBECOM2017. 1536-1276 © 2015 IEEE. Personal use is permitted, but republication/redistribution requires IEEE permission. Please find the final version in IEEE from the link: http://ieeexplore.ieee.org/document/xxxxxxx/. Digital Object Identifier: 10.1109/xxxxxxxx}
\end{abstract}

\section{Introduction\label{sec:Introduction}}

An orthogonal deployment of dense small cell networks (SCNs) within
the existing macrocell ones is characterized by small cells and macrocells
operating on different frequency spectrum. Due to its capacity gain
and easy implementation, such deployment is considered as one of the
workhorses for capacity enhancement in the 4th-generation (4G) and
5th-generation (5G) networks, developed by the 3rd Generation Partnership
Project (3GPP)~\cite{Tutor_smallcell,Ge2016UDNs}.%
{} It is also envisaged that dense SCNs will embrace time division duplexing
(TDD), which does not require a pair of frequency carriers, and offers
the possibility of tailoring the amount of time radio resources to
the downlink (DL)/uplink (UL) traffic conditions. In this line, seven
TDD configurations, each associated with a specific DL/UL subframe
number in a 10-milisecond TDD frame, are available for static or semi-static
selection at the network side in the 4G Long Term Evolution (LTE)
networks~\cite{TR36.828}. However, such static TDD operation cannot
adapt DL/UL subframe resources to the fast fluctuations in DL/UL traffic
demands.%

To overcome the drawbacks of static TDD, %
a new technology, referred to as dynamic TDD, has recently drawn much
attention~\cite{Ding2014dynTDDana,Sun2015dynTDD,Yu2015dynTDD,Gupta2016ICC_dynTDD,Ding2016dynTDD}.
In dynamic TDD, the configuration of the TDD DL/UL subframe number
in each cell or a cluster of cells can be dynamically changed on a
per-frame basis, i.e., once every 10 milliseconds~\cite{TR36.828}.
Thus, dynamic TDD can provide a tailored configuration of DL/UL subframe
resources for each cell or a cluster of cells at the expense of allowing
\emph{inter-cell inter-link interference}, e.g., DL transmissions
of a cell may interfere with UL ones of a neighboring cell, and vice
versa.

The study of dynamic TDD is particularly important because it provides
valuable insights on the full duplex (FD) transmission~\cite{Ding2016dynTDD,Goyal2016ICC_FD},
which has been identified as one of the candidate technologies for
5G. In more detail,
\begin{itemize}
\item In an FD system, a base station (BS) can simultaneously transmit to
and receive from different user equipments (UEs), thus enhancing spectrum
reuse, but creating \emph{(i) }inter-cell inter-link interference
and \emph{(ii) }intra-cell inter-link interference, a.k.a., self-interference~\cite{Goyal2016ICC_FD}.
\item The main difference between an FD system and a dynamic TDD one is
that intra-cell inter-link interference does not exist in dynamic
TDD~\cite{Ding2016dynTDD}.
\end{itemize}

The physical (PHY) layer signal-to-interference-plus-noise ratio (SINR)
performance of dynamic TDD has been analyzed in~\cite{Ding2014dynTDDana},
assuming deterministic positions of BSs and UEs, and in~\cite{Sun2015dynTDD,Yu2015dynTDD,Gupta2016ICC_dynTDD},
considering stochastic positions of BSs and UEs. The main conclusion
in~\cite{Ding2014dynTDDana,Sun2015dynTDD,Yu2015dynTDD,Gupta2016ICC_dynTDD}
was that the UL SINR of dynamic TDD suffers from severe performance
degradation due to the strong DL-to-UL interference, which is generated
from a BS to another one. To address this challenge, cell clustering
as well as full or partial interference cancellation (IC) at the BS
side were recommended in~\cite{Ding2014dynTDDhom,Ding2014dynTDDhet}%
. However, since clustering and IC are not trivial operations, such
disadvantage of dynamic TDD in terms of UL SINR triggers a fundamental
question: \textbf{What is the true value of dynamic TDD?}\textbf{\emph{
}}%
The answer can only be found through a thorough media access control
(MAC) layer analysis, since the main motivation of dynamic TDD is
to \emph{dynamically} adapt DL/UL subframe resources to the variation
of traffic demands. To the best our knowledge, the MAC layer analysis
of dynamic TDD has not been investigated from a theoretical viewpoint
in the literature, although some preliminary simulation results can
be found in~\cite{Ding2016dynTDD,Ding2014dynTDDhom,Ding2014dynTDDhet,Shen2012dynTDD}.%

In this paper, for the first time, we conduct a theoretical study
on the MAC layer performance of dynamic TDD, and present a single
contribution in this paper:%

\begin{itemize}
\item We derive closed-form expressions of \emph{the DL/UL time resource
utilization} (TRU) for synchronous dynamic TDD, which has been widely
adopted in the existing LTE systems. Such results quantify the performance
of dynamic TDD in terms of its MAC layer resource usage%
, as SCNs evolve into dense and ultra-dense ones. More specifically,
\begin{itemize}
\item We show that the DL/UL TRU varies across different TDD subframes%
.
\item We prove that the average total TRU\footnote{Later, the average total TRU will be formally defined as the sum of
the average DL TRU and the average UL TRU.} of dynamic TDD is larger than that of static TDD%
.
\item We derive the limit of the performance gain of dynamic TDD compared
with static TDD, in terms of the average total TRU%
.
\end{itemize}
\end{itemize}

\section{Network Scenario and Performance Metrics\label{sec:Network-Model}}

In this section, we present the network scenario and the performance
metrics considered in the paper.

\subsection{General Network Scenario\label{subsec:Network-Scenario}}

We consider a cellular network with BSs deployed on a plane according
to a homogeneous Poisson point process (HPPP) $\Phi$ with a density
of $\lambda$ $\textrm{BSs/km}^{2}$. Active UEs are also Poisson
distributed in the considered network with a density of $\rho$ $\textrm{UEs/km}^{2}$.
Here, we only consider active UEs in the network because non-active
UEs do not trigger data transmissions%
.%
{}

In practice, a BS will mute its transmission, if there is no UE connected
to it, which reduces inter-cell interference and energy consumption~\cite{Ding2016TWC_IMC}.
Note that such BS idle mode operation is not trivial, which even changes
the capacity scaling law~\cite{Ding2017capScaling}. Since UEs are
randomly and uniformly distributed in the network, we assume that
the active BSs also follow an HPPP distribution $\tilde{\Phi}$~\cite{dynOnOff_Huang2012},
the density of which is denoted by $\tilde{\lambda}$ $\textrm{BSs/km}^{2}$.
Note that $0\leq\tilde{\lambda}\leq\lambda$, and a larger $\rho$
leads to %
a larger $\tilde{\lambda}$%
. From~\cite{dynOnOff_Huang2012,Ding2016IMC_GC}, $\tilde{\lambda}$
is given by%
\begin{equation}
\tilde{\lambda}=\lambda\left[1-\frac{1}{\left(1+\frac{\rho}{q\lambda}\right)^{q}}\right],\label{eq:lambda_tilde_HuangKB}
\end{equation}
where $q$ takes an empirical value around 3.5\textasciitilde{}4~\cite{dynOnOff_Huang2012,Ding2016IMC_GC}\footnote{With a 3GPP path loss model incorporating both line-of-sight (LoS)
and non-LoS (NLoS) transmissions~\cite{TR36.828}, it was shown in~\cite{Ding2016TWC_IMC}
that $q=4.05$ when $\rho=300\thinspace\textrm{UEs/km}^{2}$, which
is a typical value of $\rho$ in 5G~\cite{Tutor_smallcell}. Note
that our analytical results in this paper can work with any value
of $q$.}%
.

For each active UE, the probabilities of it requesting DL and UL data
are respectively denoted by $p^{{\rm {D}}}$ and $p^{{\rm {U}}}$,
with $p^{{\rm {D}}}+p^{{\rm {U}}}=1$%
. Besides, we assume that each request is large enough to be transmitted
for at least one TDD frame, which consists of $T$ subframes. In the
sequel, the DL or UL subframe number per frame will be shortened as
the DL or UL subframe number, because subframes are meant within one
frame.

\subsection{Asynchronous and Synchronous Networks\label{subsec:two-network-cases}}

Regarding TDD frames, we need to differentiate two types of networks,
i.e., asynchronous ones and synchronous ones. Note that most previous
works only investigated dynamic TDD operating in an asynchronous network~\cite{Ding2014dynTDDana,Sun2015dynTDD,Yu2015dynTDD,Gupta2016ICC_dynTDD}.%
{} More specifically, in an asynchronous network, e.g., Wireless Fidelity
(Wi-Fi), the TDD transmission frames are not aligned among cells,
and thus both the PHY layer performance and the MAC layer one are
uniform across all TDD subframes. However, in a synchronous network,
such as LTE and the future 5G networks, the TDD transmission frames
from different cells are aligned to simplify the design of the radio
protocols. As a result, the performance of dynamic TDD becomes a function
of the TDD configuration structure.
\begin{figure}
\noindent \begin{centering}
\includegraphics[width=8cm]{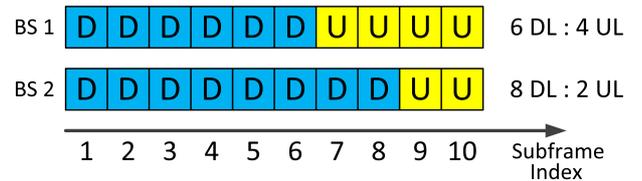}\renewcommand{\figurename}{Fig.}\caption{\label{fig:TDD_configs}An example of the LTE TDD configurations,
where one TDD frame is composed of $T=10$\,subframes. Here, 'D'
and 'U' denote a DL subframe and an UL one, respectively.}
\par\end{centering}
\vspace{-0.4cm}
\end{figure}
 Fig.~\ref{fig:TDD_configs} shows an example of such TDD configuration
structure in LTE~\cite{TR36.828}. %

In Fig.~\ref{fig:TDD_configs}, TDD frames of different cells are
aligned in the time domain. One TDD frame is composed of $T=10$\,subframes,
and the time length of each subframe is 1\,millisecond~\cite{TR36.828}.
As an example, we assume that BS\,1 and BS\,2 use the TDD configurations
with 6 and 8 DL subframes, respectively. For this synchronous network
with TDD frame alignment, we have the following two remarks.

\textbf{Remark~1: }The first few subframes are more likely to carry
DL transmissions than the last few ones. The opposite conclusion applies
for the UL. This leads to a subframe dependent MAC layer performance
of dynamic TDD.

\textbf{Remark~2: }The subframes in the middle of the frame are more
likely to be subject to inter-cell inter-link interference than those
at the two ends of the TDD frame. This implies a subframe dependent
PHY layer performance of dynamic TDD.

Unfortunately, such synchronous dynamic TDD that has been used in
LTE and will be used in 5G, has not been well treated in the literature,
and previous works considering synchronous dynamic TDD were only based
on simulations and lack of analytical rigor~\cite{Ding2016dynTDD,Ding2014dynTDDhom,Ding2014dynTDDhet,Shen2012dynTDD}.
In this paper, we explore Remark~1, and study the MAC layer performance
of synchronous dynamic TDD, with the consideration of the DL-before-UL
TDD configuration structure adopted in LTE, as illustrated in Fig.~\ref{fig:TDD_configs}%
.

\subsection{MAC Layer Performance Metrics\label{subsec:Performance-Metrics}}

For the $l$-th subframe ($l\in\left\{ 1,2,\ldots,T\right\} $), we
define the subframe dependent DL TRU and UL\emph{ }TRU as the probability
of BSs transmitting DL signals and that of UEs transmitting UL signals,
respectively, which are denoted by $q_{l}^{{\rm {D}}}$ and $q_{l}^{{\rm {U}}}$.
In other words, $q_{l}^{{\rm {D}}}$ and $q_{l}^{{\rm {U}}}$ characterize
how much time resource is actually used for DL and UL, respectively.
Note that generally we do \emph{not} have $q_{l}^{{\rm {D}}}+q_{l}^{{\rm {U}}}=1$,
because some subframe resources might be wasted, e.g., in static TDD.

Moreover, we define the average DL TRU $\kappa^{{\rm {D}}}$ and the
average UL TRU $\kappa^{{\rm {U}}}$ as the mean value of $q_{l}^{{\rm {D}}}$
and $q_{l}^{{\rm {U}}}$ across all of the $T$ subframes:
\begin{equation}
\begin{cases}
\hspace{-0.2cm}\begin{array}{l}
\kappa^{{\rm {D}}}=\frac{1}{T}\sum_{l=1}^{T}q_{l}^{{\rm {D}}}\\
\kappa^{{\rm {U}}}=\frac{1}{T}\sum_{l=1}^{T}q_{l}^{{\rm {U}}}
\end{array}\end{cases}\hspace{-0.4cm}\hspace{-0.2cm}.\label{eq:DEF_ave_DL_UL_TRU}
\end{equation}

Finally, we define the average total TRU $\kappa$ as the sum of $\kappa^{{\rm {D}}}$
and $\kappa^{{\rm {U}}}$, which is written as
\begin{equation}
\kappa=\kappa^{{\rm {D}}}+\kappa^{{\rm {U}}}.\label{eq:DEF_ave_tot_TRU}
\end{equation}

In the following sections, we will investigate the performance of
$q_{l}^{Link}$, $\kappa^{Link}$ and $\kappa$ considering Remark~1,
where the string variable $Link$ denotes the link direction, and
takes the value of 'D' and 'U' for the DL and the UL respectively\footnote{On a side node, the DL/UL area spectral efficiency (ASE) in $\textrm{bps/Hz/km}^{2}$
can be further computed by $\frac{\tilde{\lambda}}{T}\sum_{l=1}^{T}q_{l}^{Link}r_{l}^{Link}$,
where $r_{l}^{Link}$ is the PHY layer per-BS data rate in $\textrm{bps/Hz/BS}$
for subframe $l$. As explained in Remark~2, the derivation of $r_{l}^{Link}$
should consider the subframe dependent inter-cell inter-link interference,
which is out of the scope of this paper since our focus is on the
MAC layer performance. The study on $r_{l}^{Link}$ will be left as
our future work. }.%

\section{Main Results\label{sec:Main-Results}}

The main goal of this section is to derive theoretical results on
the DL/UL TRU defined in Subsection~\ref{subsec:Performance-Metrics}.%
{} Note that computing $q_{l}^{{\rm {D}}}$ and $q_{l}^{{\rm {U}}}$
is a non-trivial task for synchronous dynamic TDD, because it involves
the following distributions:
\begin{itemize}
\item the distribution of the UE number in an active BS, which will be shown
to follow \textbf{a truncated Negative Binomial distribution} in Subsection~\ref{subsec:PMF_actUE_num};
\item the distribution of the DL/UL data request numbers in an active BS,
which will be shown to follow \textbf{a Binomial distribution} in
Subsection~\ref{subsec:PMF_DL_dataReq_num};
\item the dynamic TDD subframe splitting strategy and the corresponding
distribution of the DL/UL subframe number, which will be shown to
follow \textbf{an aggregated Binomial distribution} in Subsection~\ref{subsec:PMF_DL_subf_num};
and finally
\item the prior information about the TDD frame structure, such as the DL-before-UL
structure adopted in LTE~\cite{TR36.828} (see Fig.~\ref{fig:TDD_configs}),
which will lead to subframe dependent results of $q_{l}^{{\rm {D}}}$
and $q_{l}^{{\rm {U}}}$ to be presented in Subsection~\ref{subsec:Pr_DL_UL_Tx_per_subf}.
\end{itemize}

\subsection{The Distribution of the UE Number in an Active BS: A Truncated Negative
Binomial Distribution\label{subsec:PMF_actUE_num}}

According to~\cite{dynOnOff_Huang2012}, %
the coverage area size $X$ of a BS can be approximately characterized
by a Gamma distribution~\cite{dynOnOff_Huang2012}, and its probability
density function (PDF) is written as
\begin{equation}
f_{X}(x)=\left(q\lambda\right)^{q}x^{q-1}\frac{\exp(\text{\textminus}q\lambda x)}{\text{\ensuremath{\Gamma}}(q)},\label{eq:cov_area_size_PDF_Gamma}
\end{equation}
where %
$\text{\ensuremath{\Gamma}}(\cdot)$ is the Gamma function~\cite{Book_Integrals}.
Then, the UE number per BS can be denoted by a random variable (RV)
$K$, and the probability mass function (PMF) of $K$ can be derived
as
\begin{eqnarray}
f_{K}\left(k\right)\hspace{-0.2cm} & = & \hspace{-0.2cm}\Pr\left[K=k\right]\nonumber \\
\hspace{-0.2cm} & \overset{\left(a\right)}{=} & \hspace{-0.2cm}\int_{0}^{+\infty}\frac{\left(\rho x\right)^{k}}{k!}\exp(-\rho x)f_{X}(x)dx\nonumber \\
\hspace{-0.2cm} & \overset{\left(b\right)}{=} & \hspace{-0.2cm}\frac{\Gamma(k+q)}{\Gamma(k+1)\Gamma(q)}\left(\frac{\rho}{\rho+q\lambda}\right)^{k}\left(\frac{q\lambda}{\rho+q\lambda}\right)^{q},\label{eq:NB_PMF}
\end{eqnarray}
where $(a)$ is due to the HPPP distribution of UEs and $(b)$ is
obtained from (\ref{eq:cov_area_size_PDF_Gamma}).%
{} It can be seen from (\ref{eq:NB_PMF}) that $K$ follows a Negative
Binomial distribution~\cite{Book_Integrals}, i.e., $K\sim\textrm{NB}\left(q,\frac{\rho}{\rho+q\lambda}\right)$.

As discussed in Subsection~\ref{subsec:Network-Scenario}, we assume
that a BS with $K=0$ is not active, which will be ignored in our
analysis due to its muted transmission. Hence, we focus on the active
BSs and further study \emph{the distribution of the UE number per
active BS}, which is denoted by a positive RV $\tilde{K}$. Considering
(\ref{eq:NB_PMF}) and the fact that the only difference between $K$
and $\tilde{K}$ lies in $\tilde{K}\neq0$, we can conclude that $\tilde{K}$
follows a truncated Negative Binomial distribution, i.e., $\tilde{K}\sim\textrm{truncNB}\left(q,\frac{\rho}{\rho+q\lambda}\right)$.
More specifically, the PMF of $\tilde{K}$ is denoted by $f_{\tilde{K}}\left(\tilde{k}\right),\tilde{k}\in\left\{ 1,2,\ldots,+\infty\right\} $,
and it is given by%
\textbf{}%
\begin{equation}
f_{\tilde{K}}\left(\tilde{k}\right)=\Pr\left[\tilde{K}=\tilde{k}\right]=\frac{f_{K}\left(\tilde{k}\right)}{1-f_{K}\left(0\right)},\label{eq:truncNB_PMF}
\end{equation}
where the denominator $\left(1-f_{K}\left(0\right)\right)$ represents
the probability of a BS being active, i.e., $\tilde{\lambda}/\lambda$.

\begin{figure}
\noindent \begin{centering}
\includegraphics[width=8cm]{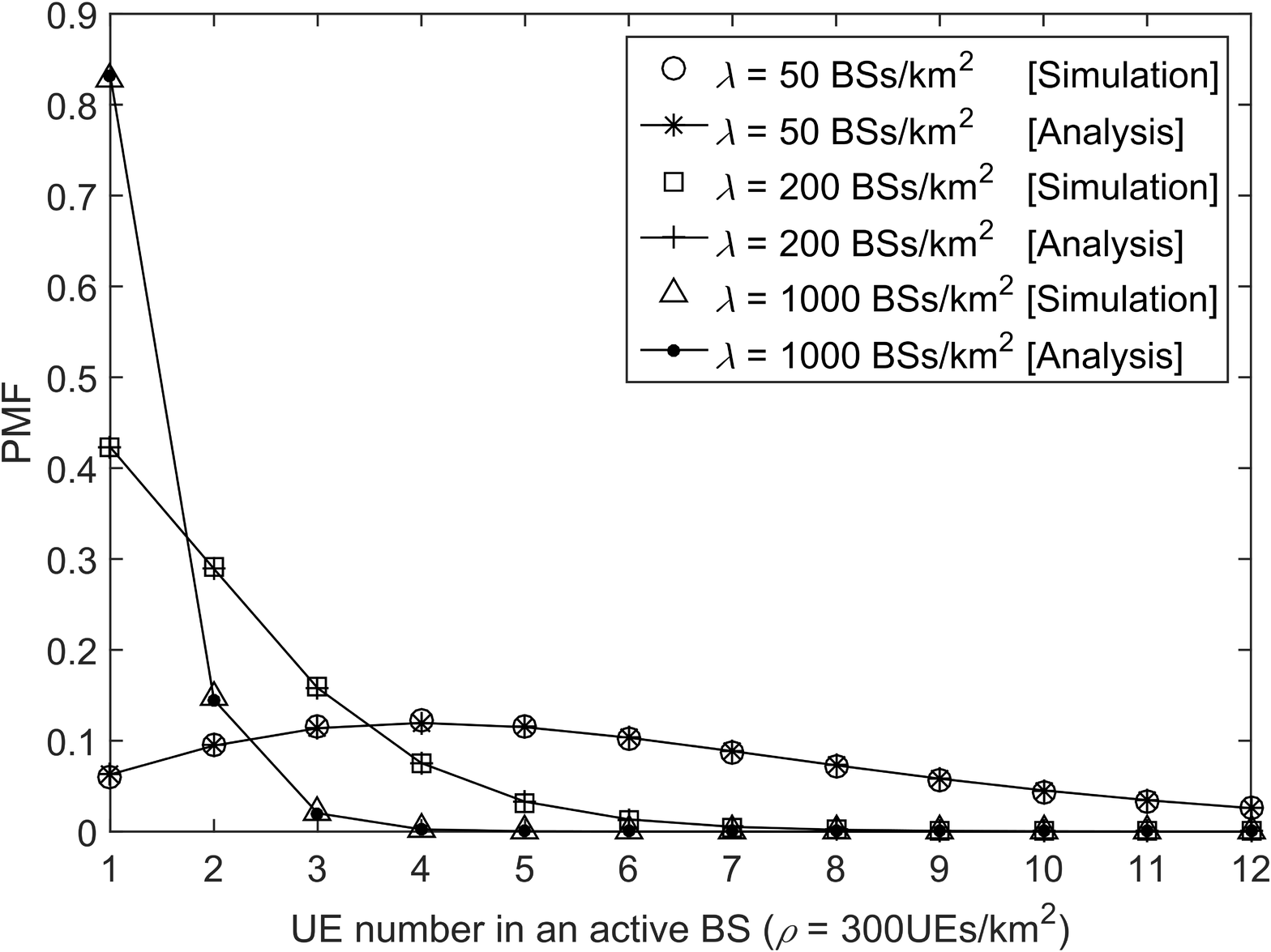}\renewcommand{\figurename}{Fig.}\caption{\label{fig:PMF_UE_num_per_actBS_rho600_various_lambda}The PMF of
the UE number $\tilde{K}$ in an active BS ($\rho=300\thinspace\textrm{UEs/km}^{2}$,
$q=4.05$, various $\lambda$).}
\par\end{centering}
\vspace{-0.4cm}
\end{figure}

\textbf{Example:} As an example, we plot the results of $f_{\tilde{K}}\left(\tilde{k}\right)$
in Fig.~\ref{fig:PMF_UE_num_per_actBS_rho600_various_lambda} with
the following parameter values: UE density $\rho=300\thinspace\textrm{UEs/km}^{2}$,
BS density $\lambda\in\left\{ 50,200,1000\right\} \thinspace\textrm{BSs/km}^{2}$
and $q=4.05$~\cite{Ding2016TWC_IMC}.

From this figure, we can draw the following observations:
\begin{itemize}
\item The analytical results based on the truncated Negative Binomial distribution
match well with the simulation results. More specifically, the maximum
difference between the simulated PMF and the analytical PMF is shown
to be less than 0.5\ percentile.
\item For the active BSs, $f_{\tilde{K}}\left(\tilde{k}\right)$ shows a
more dominant peak at $\tilde{k}=1$ as $\lambda$, and in turn $\tilde{\lambda}$,
increases. This is because the ratio of $\rho$ to $\tilde{\lambda}$
gradually decreases toward 1 as $\lambda$ increases, approaching
the limit of one UE per active BS in ultra-dense SCNs. In particular,
when $\lambda=1000\thinspace\textrm{BSs/km}^{2}$, more than 80$\thinspace$\%
of the active BSs will serve only one UE. Intuitively speaking, each
of those BSs should dynamically engage all subframes for DL or UL
transmissions based on the specific traffic demand of its served UE.
In such way, the subframe resources can be fully utilized in dynamic
TDD. Note that such operation is not available in static TDD due to
its fixed TDD configuration.
\end{itemize}

\subsection{The Distribution of the DL/UL Data Request Number in an Active BS:
A Binomial Distribution\label{subsec:PMF_DL_dataReq_num}}

After obtaining $f_{\tilde{K}}\left(\tilde{k}\right)$%
, we need to further study the distribution of the DL/UL data request
number in an active BS, so that a tailored TDD configuration can be
determined in a dynamic TDD network. %

For clarity, the DL and UL data request numbers in an active BS are
denoted by RVs $M^{{\rm {D}}}$ and $M^{{\rm {U}}}$, respectively.
Since we assume that each UE generates one request of either DL data
or UL data (see Subsection~\ref{subsec:Network-Scenario}), it is
easy to show that
\begin{equation}
M^{{\rm {D}}}+M^{{\rm {U}}}=\tilde{K}.\label{eq:duality_MD_MU}
\end{equation}
As discussed in Subsection~\ref{subsec:Network-Scenario}, for each
UE in an active BS, the probability of it requesting DL data and UL
data is $p^{{\rm {D}}}$ and $p^{{\rm {U}}}$, respectively. Hence,
for a given UE number $\tilde{k}$, $M^{{\rm {D}}}$ and $M^{{\rm {U}}}$
follow Binomial distributions~\cite{Book_Integrals}, i.e., $M^{{\rm {D}}}\sim\textrm{Bi}\left(\tilde{k},p^{{\rm {D}}}\right)$
and $M^{{\rm {U}}}\sim\textrm{Bi}\left(\tilde{k},p^{{\rm {U}}}\right)$.
More specifically, the PMFs of $M^{{\rm {D}}}$ and $M^{{\rm {U}}}$
can be respectively written as
\begin{eqnarray}
f_{M^{{\rm {D}}}}\left(m^{{\rm {D}}}\right)\hspace{-0.2cm} & = & \hspace{-0.2cm}\left(\begin{array}{c}
\tilde{k}\\
m^{{\rm {D}}}
\end{array}\right)\left(p^{{\rm {D}}}\right)^{m^{{\rm {D}}}}\left(1-p^{{\rm {D}}}\right)^{\tilde{k}-m^{{\rm {D}}}},\label{eq:DL_dataReq_PMF_Bino}
\end{eqnarray}
and%
\begin{eqnarray}
f_{M^{{\rm {U}}}}\left(m^{{\rm {U}}}\right)\hspace{-0.2cm} & = & \hspace{-0.2cm}\left(\begin{array}{c}
\tilde{k}\\
m^{{\rm {U}}}
\end{array}\right)\left(p^{{\rm {U}}}\right)^{m^{{\rm {U}}}}\left(1-p^{{\rm {U}}}\right)^{\tilde{k}-m^{{\rm {U}}}}.\label{eq:UL_dataReq_PMF_Bino}
\end{eqnarray}

\begin{figure}
\noindent \begin{centering}
\includegraphics[width=8cm]{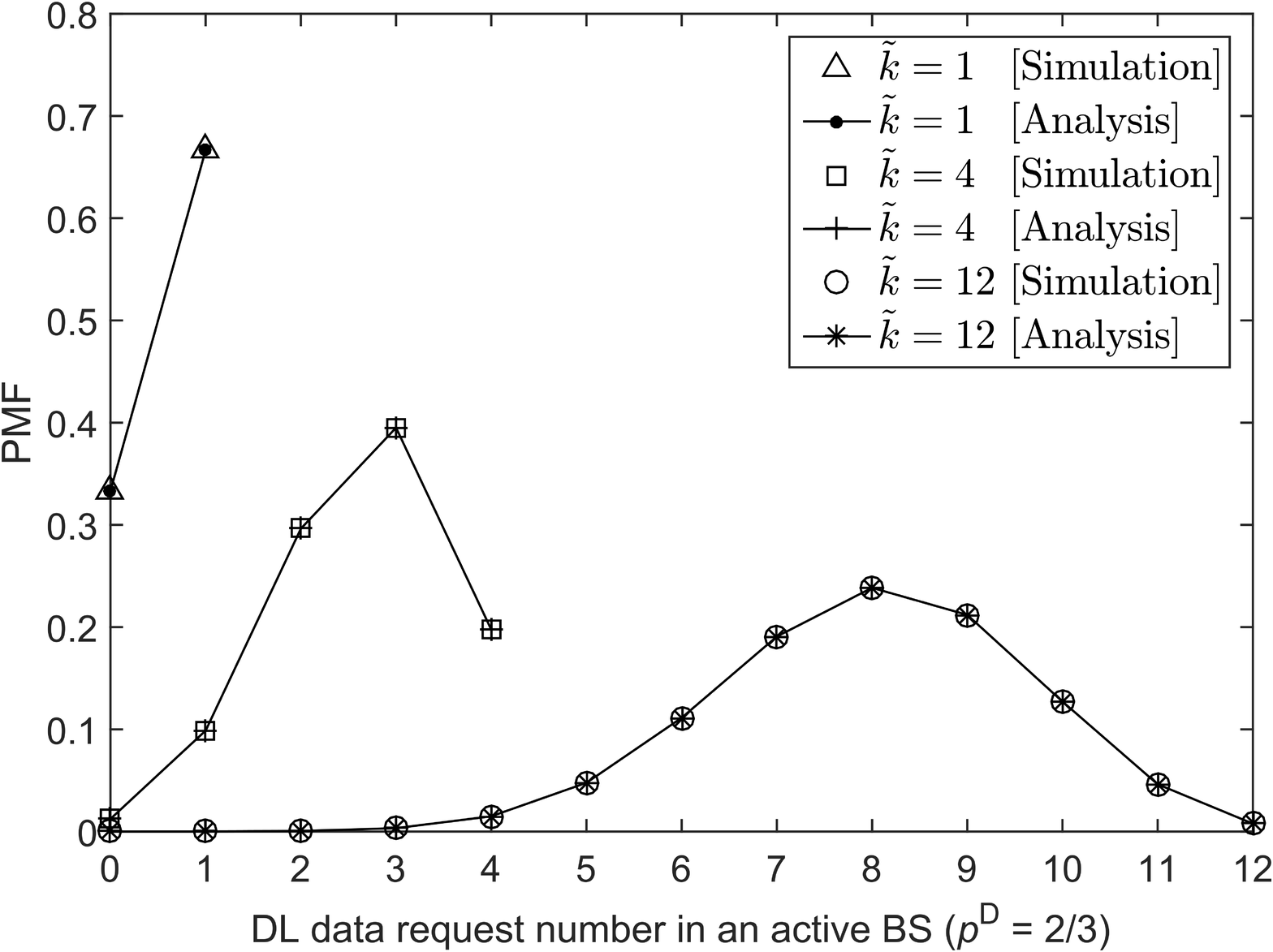}\renewcommand{\figurename}{Fig.}\caption{\label{fig:PMF_DL_dataReq_num_pD_0p66_various_k_tilde}The PMF of
the DL data request number $M^{{\rm {D}}}$ in an active BS ($p^{{\rm {D}}}=\frac{2}{3}$,
various $\tilde{k}$).}
\par\end{centering}
\vspace{-0.4cm}
\end{figure}

\textbf{Example:} As an example, we plot the results of $f_{M^{{\rm {D}}}}\left(m^{{\rm {D}}}\right)$
in Fig.~\ref{fig:PMF_DL_dataReq_num_pD_0p66_various_k_tilde} with
the following parameter values: $p^{{\rm {D}}}=\frac{2}{3}$ and $\tilde{k}\in\left\{ 1,4,12\right\} $.
For brevity, we omit displaying the PMF of $M^{{\rm {U}}}$, due to
its duality to $M^{{\rm {D}}}$ as shown in (\ref{eq:duality_MD_MU}).

From this figure, we can draw the following observations:
\begin{itemize}
\item The Binomial distribution accurately characterizes the distribution
of $M^{{\rm {D}}}$, as the analytical PMFs well match the simulated
ones.
\item The average value of $M^{{\rm {D}}}$ is around $p^{{\rm {D}}}\tilde{k}$,
which is inline with intuition.
\end{itemize}

\subsection{The Distribution of the DL/UL Subframe Number with Dynamic TDD: An
Aggregated Binomial Distribution\label{subsec:PMF_DL_subf_num}}

After knowing the distribution of the DL data request number $M^{{\rm {D}}}$
in an active BS, we are now ready to consider dynamic TDD, and derive
the distribution of the DL subframe number in an active BS. For a
given UE number $\tilde{k}$, the DL subframe number in an active
BS is denoted by $N^{{\rm {D}}}$. Here, we adopt a dynamic TDD algorithm
to choose the DL subframe number, which matches the DL subframe ratio
with the DL data request ratio~\cite{Ding2016dynTDD}. In more detail,
for certain values of $m^{{\rm {D}}}$ and $\tilde{k}$, the DL subframe
number $n\left(m^{{\rm {D}}},\tilde{k}\right)$ is determined by
\begin{equation}
n\left(m^{{\rm {D}}},\tilde{k}\right)=\textrm{round}\left(\frac{m^{{\rm {D}}}}{\tilde{k}}T\right),\label{eq:DL_subf_selection_criterion}
\end{equation}
where $\textrm{round}\left(x\right)$ is an operator that rounds a
real value $x$ to its nearest integer. In (\ref{eq:DL_subf_selection_criterion}),
$\frac{m^{{\rm {D}}}}{\tilde{k}}$ can be deemed as the DL data request
ratio, because \emph{(i)} $m^{{\rm {D}}}$ denotes the DL data request
number with its distribution characterized in (\ref{eq:DL_dataReq_PMF_Bino});
and \emph{(ii)} $\tilde{k}$ represents the UE number, and thus the
total number of the DL and UL data requests. As a result, $\frac{m^{{\rm {D}}}}{\tilde{k}}T$
yields the desirable DL subframe number that matches the DL subframe
ratio with the DL data request ratio. Due to the integer nature of
the DL subframe number, we use the round operator to generate a valid
DL subframe number that is nearest to $\frac{m^{{\rm {D}}}}{\tilde{k}}T$
in (\ref{eq:DL_subf_selection_criterion}).

Based on (\ref{eq:DL_subf_selection_criterion}), the PMF of $N^{{\rm {D}}}$
is denoted by $f_{N^{{\rm {D}}}}\left(n^{{\rm {D}}}\right),n^{{\rm {D}}}\in\left\{ 0,1,\ldots,T\right\} $,
and it can be derived as
\begin{eqnarray}
\hspace{-0.4cm}\hspace{-0.4cm}f_{N^{{\rm {D}}}}\left(n^{{\rm {D}}}\right)\hspace{-0.3cm} & = & \hspace{-0.3cm}\Pr\left[N^{{\rm {D}}}=n^{{\rm {D}}}\right]\nonumber \\
\hspace{-0.3cm} & \overset{\left(a\right)}{=} & \hspace{-0.3cm}\hspace{-0.1cm}\sum_{m^{{\rm {D}}}=0}^{\tilde{k}}\hspace{-0.1cm}I\hspace{-0.1cm}\left\{ \textrm{round}\left(\frac{m^{{\rm {D}}}}{\tilde{k}}T\right)\hspace{-0.1cm}=\hspace{-0.1cm}n^{{\rm {D}}}\right\} \hspace{-0.1cm}f_{M^{{\rm {D}}}}\left(m^{{\rm {D}}}\right)\hspace{-0.1cm},\label{eq:DL_subf_PMF}
\end{eqnarray}
where (\ref{eq:DL_subf_selection_criterion}) is plugged into (a),
and $I\left\{ X\right\} $ is an indicator function that outputs one
when $X$ is true and zero otherwise. Besides, $f_{M^{{\rm {D}}}}\left(m^{{\rm {D}}}\right)$
is computed by (\ref{eq:DL_dataReq_PMF_Bino}). Due to the existence
of the indicator function in (\ref{eq:DL_subf_PMF}), $f_{N^{{\rm {D}}}}\left(n^{{\rm {D}}}\right)$
can be viewed as an aggregated PMF of Binomial PMFs, since $N^{{\rm {D}}}$
is computed from $M^{{\rm {D}}}$ according to a many-to-one mapping
in (\ref{eq:DL_subf_selection_criterion}). %

Because the total subframe number in a frame is $T$, and each subframe
should be either a DL one or an UL one, it is apparent that $N^{{\rm {D}}}+N^{{\rm {U}}}=T$,
and thus we have
\begin{equation}
f_{N^{{\rm {U}}}}\left(n^{{\rm {U}}}\right)=f_{N^{{\rm {D}}}}\left(T-n^{{\rm {U}}}\right).\label{eq:UL_subf_PMF}
\end{equation}

\begin{figure}
\noindent \begin{centering}
\includegraphics[width=8cm]{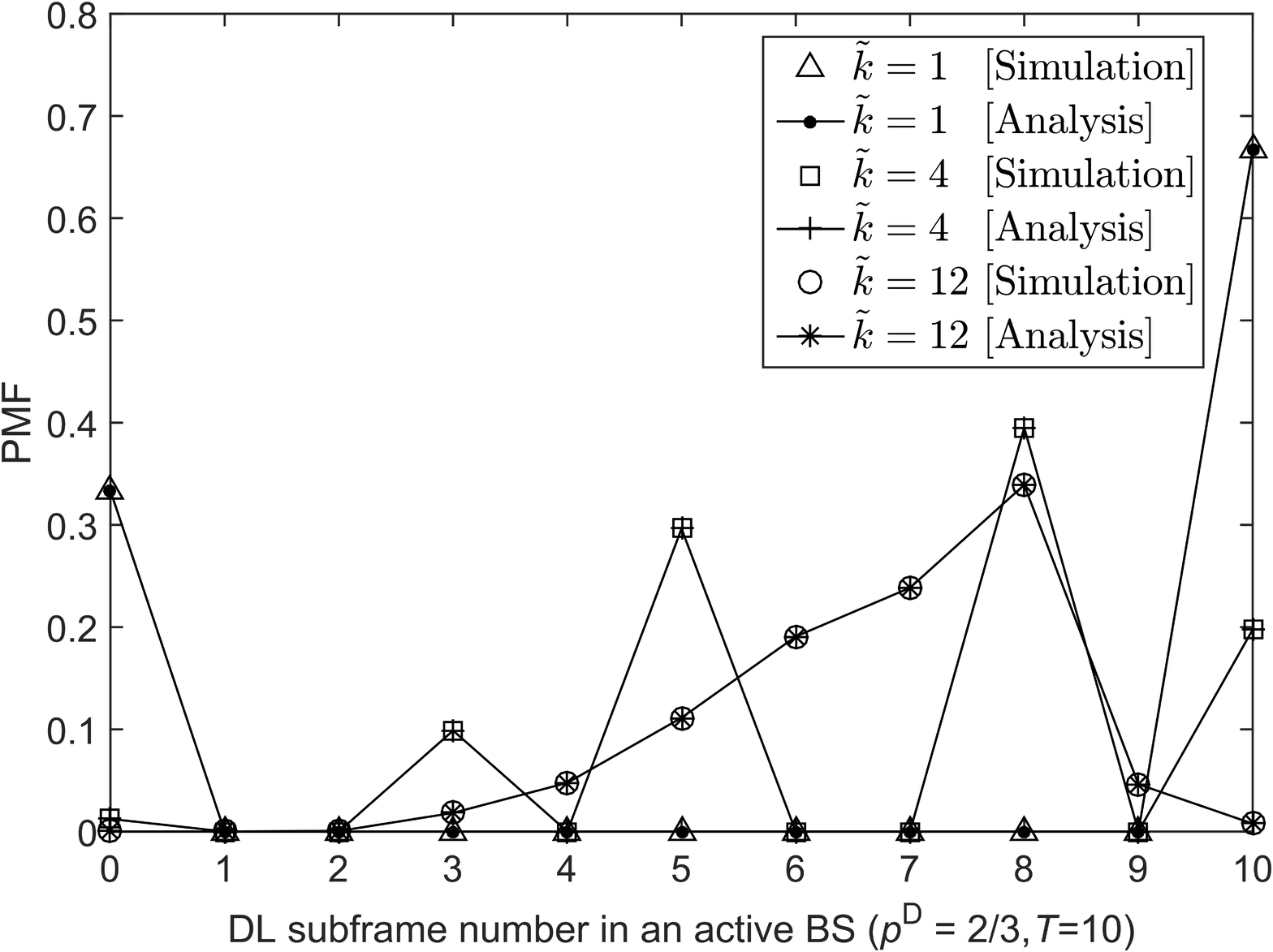}\renewcommand{\figurename}{Fig.}\caption{\label{fig:PMF_DL_subf_num_pD_0p66_various_k_tilde}The PMF of the
DL subframe number $N^{{\rm {D}}}$ in an active BS ($p^{{\rm {D}}}=\frac{2}{3}$,
$T=10$, various $\tilde{k}$).}
\par\end{centering}
\vspace{-0.4cm}
\end{figure}

\textbf{Example:} As an example, we plot the results of $f_{N^{{\rm {D}}}}\left(n^{{\rm {D}}}\right)$
in Fig.~\ref{fig:PMF_DL_subf_num_pD_0p66_various_k_tilde} with the
following parameter values: $p^{{\rm {D}}}=\frac{2}{3}$, $T=10$
and $\tilde{k}\in\left\{ 1,4,12\right\} $.

From this figure, we can draw the following observations:
\begin{itemize}
\item Following the logic of dynamic TDD characterized in (\ref{eq:DL_subf_selection_criterion}),
when $\tilde{k}=1$, the DL subframe number is set to either $T=10$
or 0, meaning that the active BS dynamically invest all subframes
in DL transmissions or UL transmissions based on the instantaneous
data request being DL (with a probability of $p^{{\rm {D}}}$) or
UL (with a probability of $p^{{\rm {U}}}$). Such strategy fully utilizes
the subframe resources in a dynamic TDD network.
\item When $\tilde{k}>1$, the PMF of the DL subframe number turns out to
be rather complex, because multiple DL data request numbers may be
mapped to the same DL subframe number due to the dynamic TDD algorithm
given by (\ref{eq:DL_subf_selection_criterion}). For example, when
$\tilde{k}=12$, we have $f_{N^{{\rm {D}}}}\left(8\right)=0.339$
as exhibited in Fig.~\ref{fig:PMF_DL_subf_num_pD_0p66_various_k_tilde}.
This is because:
\begin{itemize}
\item according to (\ref{eq:DL_subf_selection_criterion}), both $M^{{\rm {D}}}=9$
and $M^{{\rm {D}}}=10$ result in $N^{{\rm {D}}}=8$, since $\textrm{round}\left(\frac{9}{12}\times10\right)=8$
and $\textrm{round}\left(\frac{10}{12}\times10\right)=8$. In other
words, with dynamic TDD, 8 subframes will be allocated for the DL,
if either 9 or 10 data requests in $\tilde{k}=12$ data requests are
DL;
\item as shown in Fig.~\ref{fig:PMF_DL_dataReq_num_pD_0p66_various_k_tilde},
$f_{M^{{\rm {D}}}}\left(9\right)=0.212$ and $f_{M^{{\rm {D}}}}\left(10\right)=0.127$;
and thus
\item from (\ref{eq:DL_subf_PMF}), we have $f_{N^{{\rm {D}}}}\left(8\right)=f_{M^{{\rm {D}}}}\left(9\right)+f_{M^{{\rm {D}}}}\left(10\right)=0.339$.
In other words, with dynamic TDD, the probability of allocating 8
DL subframes equals to the sum of the probabilities of observing 9
and 10 DL data requests in 12 data requests.
\end{itemize}
\end{itemize}

\subsection{The Subframe Dependent DL/UL TRU\label{subsec:Pr_DL_UL_Tx_per_subf}}

In this subsection, we present our main results on the subframe dependent
DL/UL TRUs $q_{l}^{{\rm {D}}}$ and $q_{l}^{{\rm {U}}}$ for dynamic
TDD in Theorem~\ref{thm:subf_depend_TRU_dynTDD}.%
\textbf{}%

\begin{thm}
\label{thm:subf_depend_TRU_dynTDD}For dynamic TDD, $q_{l}^{{\rm {D}}}$
and $q_{l}^{{\rm {U}}}$ are given by
\begin{equation}
\begin{cases}
\begin{array}{l}
q_{l}^{{\rm {D}}}=\sum_{\tilde{k}=1}^{+\infty}\left(1-\sum_{i=0}^{l-1}f_{N^{{\rm {D}}}}\left(i\right)\right)f_{\tilde{K}}\left(\tilde{k}\right)\\
q_{l}^{{\rm {U}}}=\sum_{\tilde{k}=1}^{+\infty}\sum_{i=0}^{l-1}f_{N^{{\rm {D}}}}\left(i\right)f_{\tilde{K}}\left(\tilde{k}\right)
\end{array}\end{cases}\hspace{-0.4cm},\label{eq:subf_l_is_DL_UL}
\end{equation}
where $f_{N^{{\rm {D}}}}\hspace{-0.1cm}\left(i\right)$ and $f_{\tilde{K}}\hspace{-0.1cm}\left(\tilde{k}\right)$
are given by (\ref{eq:DL_subf_PMF}) and (\ref{eq:truncNB_PMF}),
respectively.
\end{thm}
\begin{IEEEproof}
Based on (\ref{eq:DL_subf_PMF}) and conditioned on $\tilde{K}=\tilde{k}$,
the probability of performing a DL transmission in subframe $l$ $\left(l\in\left\{ 1,2,\ldots,T\right\} \right)$
can be calculated as
\begin{eqnarray}
q_{l,\tilde{k}}^{{\rm {D}}} & = & \Pr\left[\left.Y_{l}={\rm {\textrm{'}D\textrm{'}}}\right|\tilde{K}=\tilde{k}\right]\nonumber \\
 & \overset{\left(a\right)}{=} & \Pr\left[N^{{\rm {D}}}\geq l\right]\nonumber \\
 & = & 1-F_{N^{{\rm {D}}}}\left(l-1\right),\label{eq:condPr_subf_l_is_DL}
\end{eqnarray}
where $Y_{l}$ denotes the link direction for the transmission on
the $l$-th subframe, which takes a string value of 'D' and 'U' for
the DL and the UL, respectively. Besides, the step $(a)$ of (\ref{eq:condPr_subf_l_is_DL})
is due to the LTE TDD configuration structure shown in Fig.~\ref{fig:TDD_configs},
and $F_{N^{{\rm {D}}}}\left(n^{{\rm {D}}}\right)$ is the cumulative
mass function (CMF) of $N^{{\rm {D}}}$ in an active BS, which is
written as
\begin{equation}
F_{N^{{\rm {D}}}}\left(n^{{\rm {D}}}\right)=\Pr\left[N^{{\rm {D}}}\leq n^{{\rm {D}}}\right]=\sum_{i=0}^{n^{{\rm {D}}}}f_{N^{{\rm {D}}}}\left(i\right).\label{eq:DL_subf_CMF}
\end{equation}
Considering the dynamic allocation of subframe to the DL and the UL
in dynamic TDD, the conditional probability of performing an UL transmission
in subframe $l$ is given by
\begin{equation}
q_{l,\tilde{k}}^{{\rm {U}}}=1-q_{l,\tilde{k}}^{{\rm {D}}}=F_{N^{{\rm {D}}}}\left(l-1\right).\label{eq:condPr_subf_l_is_UL}
\end{equation}

Furthermore, the unconditional probabilities of performing a DL and
UL transmissions on the $l$-th subframe, i.e., $q_{l}^{{\rm {D}}}$
and $q_{l}^{{\rm {U}}}$, can be respectively derived by calculating
the expected values of $q_{l,\tilde{k}}^{{\rm {D}}}$ and $q_{l,\tilde{k}}^{{\rm {U}}}$
over all the possible values of $\tilde{k}$ as shown in (\ref{eq:subf_l_is_DL_UL}),
which concludes our proof.
\end{IEEEproof}

\subsection{The Average DL/UL/Total TRU\label{subsec:TRU_dynTDD}}

From Theorem~\ref{thm:subf_depend_TRU_dynTDD}, and Equations~(\ref{eq:DEF_ave_DL_UL_TRU})
and (\ref{eq:DEF_ave_tot_TRU}), we can obtain the results on the
average DL/UL/total TRU for dynamic TDD, which are summarized in Lemma~\ref{lem:ave_TRU_dynTDD}.
\begin{lem}
\label{lem:ave_TRU_dynTDD}For dynamic TDD, $\left\{ \kappa^{{\rm {D}}},\kappa^{{\rm {U}}},\kappa\right\} $
is given by
\begin{equation}
\begin{cases}
\hspace{-0.2cm}\begin{array}{l}
\kappa^{{\rm {D}}}=\frac{1}{T}\sum_{l=1}^{T}\sum_{\tilde{k}=1}^{+\infty}\left(1-\sum_{i=0}^{l-1}f_{N^{{\rm {D}}}}\left(i\right)\right)f_{\tilde{K}}\left(\tilde{k}\right)\\
\kappa^{{\rm {U}}}=\frac{1}{T}\sum_{l=1}^{T}\sum_{\tilde{k}=1}^{+\infty}\sum_{i=0}^{l-1}f_{N^{{\rm {D}}}}\left(i\right)f_{\tilde{K}}\left(\tilde{k}\right)\\
\kappa=1
\end{array}\end{cases}\hspace{-0.4cm}\hspace{-0.2cm},\label{eq:Pr_DL_UL_Tx_dynTDD}
\end{equation}
where $f_{N^{{\rm {D}}}}\hspace{-0.1cm}\left(i\right)$ and $f_{\tilde{K}}\hspace{-0.1cm}\left(\tilde{k}\right)$
are given by (\ref{eq:DL_subf_PMF}) and (\ref{eq:truncNB_PMF}),
respectively.
\end{lem}
\begin{IEEEproof}
The proof is straightforward by plugging (\ref{eq:subf_l_is_DL_UL})
into (\ref{eq:DEF_ave_DL_UL_TRU}) and (\ref{eq:DEF_ave_tot_TRU}).
\end{IEEEproof}

Lemma~\ref{lem:ave_TRU_dynTDD} not only quantifies the average MAC
layer performance of dynamic TDD, but also shows from a theoretical
viewpoint that dynamic TDD always achieves a full resource utilization,
i.e., $\kappa=1$, thanks to the smart adaption of DL/UL subframes
to DL/UL data requests.%

Next, we present our main results on $\kappa^{{\rm {D}}}$, $\kappa^{{\rm {U}}}$
and $\kappa$ for static TDD in Theorem~\ref{thm:ave_TRU_statTDD}.
\begin{thm}
\label{thm:ave_TRU_statTDD}For static TDD, $\left\{ \kappa^{{\rm {D}}},\kappa^{{\rm {U}}},\kappa\right\} $
can be derived as
\begin{equation}
\begin{cases}
\hspace{-0.2cm}\begin{array}{l}
\kappa^{{\rm {D}}}=\left(1-\sum_{\tilde{k}=1}^{+\infty}\left(1-p^{{\rm {D}}}\right)^{\tilde{k}}f_{\tilde{K}}\left(\tilde{k}\right)\right)\frac{N_{0}^{{\rm {D}}}}{T}\\
\kappa^{{\rm {U}}}=\left(1-\sum_{\tilde{k}=1}^{+\infty}\left(1-p^{{\rm {U}}}\right)^{\tilde{k}}f_{\tilde{K}}\left(\tilde{k}\right)\right)\frac{N_{0}^{{\rm {U}}}}{T}\\
\kappa\hspace{-0.1cm}=\hspace{-0.1cm}\frac{1}{T}\hspace{-0.1cm}\sum_{\tilde{k}=1}^{+\infty}\hspace{-0.1cm}\left[\left(\hspace{-0.1cm}1-\left(p^{{\rm {U}}}\right)^{\tilde{k}}\right)\hspace{-0.1cm}N_{0}^{{\rm {D}}}\hspace{-0.1cm}+\hspace{-0.1cm}\left(\hspace{-0.1cm}1-\left(p^{{\rm {D}}}\right)^{\tilde{k}}\right)\hspace{-0.1cm}N_{0}^{{\rm {U}}}\right]\hspace{-0.1cm}f_{\tilde{K}}\hspace{-0.1cm}\left(\tilde{k}\right)
\end{array}\end{cases}\hspace{-0.4cm}\hspace{-0.1cm}\hspace{-0.1cm}\hspace{-0.1cm}\hspace{-0.1cm}\hspace{-0.1cm}\hspace{-0.1cm}\hspace{-0.1cm}\hspace{-0.1cm},\label{eq:Pr_DL_UL_Tx_statTDD}
\end{equation}
where $f_{\tilde{K}}\left(\tilde{k}\right)$ is obtained from (\ref{eq:truncNB_PMF}),
and $N_{0}^{{\rm {D}}}$ and $N_{0}^{{\rm {U}}}$ are the designated
subframe numbers for the DL and the UL in static TDD, respectively,
which satisfy $N_{0}^{{\rm {D}}}+N_{0}^{{\rm {U}}}=T$.
\end{thm}
\begin{IEEEproof}
In static TDD, for a given UE number $\tilde{k}$ in an active BS,
the probabilities that no UE requests any DL data and no UE requests
any UL data can be calculated by $f_{M^{{\rm {D}}}}\left(0\right)$
from (\ref{eq:DL_dataReq_PMF_Bino}) and $f_{M^{{\rm {U}}}}\left(0\right)$
from (\ref{eq:UL_dataReq_PMF_Bino}), respectively. Even in such cases,
static TDD still unwisely allocates $N_{0}^{{\rm {D}}}$ and $N_{0}^{{\rm {U}}}$
subframes for the DL and the UL, respectively, which causes resource
waste. The probabilities of such resource waste for the DL and the
UL are denoted by $w^{{\rm {D}}}$ and $w^{{\rm {U}}}$, and they
can be calculated as
\begin{equation}
\begin{cases}
\hspace{-0.2cm}\begin{array}{l}
w^{{\rm {D}}}=\sum_{\tilde{k}=1}^{+\infty}f_{M^{{\rm {D}}}}\left(0\right)f_{\tilde{K}}\hspace{-0.1cm}\left(\tilde{k}\right)=\sum_{\tilde{k}=1}^{+\infty}\left(1-p^{{\rm {D}}}\right)^{\tilde{k}}\hspace{-0.1cm}f_{\tilde{K}}\hspace{-0.1cm}\left(\tilde{k}\right)\\
w^{{\rm {U}}}=\sum_{\tilde{k}=1}^{+\infty}f_{M^{{\rm {U}}}}\left(0\right)f_{\tilde{K}}\hspace{-0.1cm}\left(\tilde{k}\right)=\sum_{\tilde{k}=1}^{+\infty}\left(1-p^{{\rm {U}}}\right)^{\tilde{k}}\hspace{-0.1cm}f_{\tilde{K}}\hspace{-0.1cm}\left(\tilde{k}\right)
\end{array}\end{cases}\hspace{-0.4cm}.\label{eq:Pr_DL_UL_resource_waste_statTDD}
\end{equation}
Excluding such resource waste from $N_{0}^{{\rm {D}}}$ and $N_{0}^{{\rm {U}}}$,
we can obtain $\kappa^{{\rm {D}}}$ and $\kappa^{{\rm {U}}}$ for
static TDD as
\begin{equation}
\begin{cases}
\hspace{-0.2cm}\begin{array}{l}
\kappa^{{\rm {D}}}=\left(1-w^{{\rm {D}}}\right)\frac{N_{0}^{{\rm {D}}}}{T}\\
\kappa^{{\rm {U}}}=\left(1-w^{{\rm {U}}}\right)\frac{N_{0}^{{\rm {U}}}}{T}
\end{array}\end{cases}\hspace{-0.4cm}.\label{eq:Pr_DL_UL_Tx_statTDD_proof}
\end{equation}
Our proof is thus completed by plugging (\ref{eq:Pr_DL_UL_resource_waste_statTDD}),
(\ref{eq:DL_dataReq_PMF_Bino}) and (\ref{eq:UL_dataReq_PMF_Bino})
into (\ref{eq:Pr_DL_UL_Tx_statTDD_proof}), followed by computing
$\kappa$ from (\ref{eq:DEF_ave_tot_TRU}).
\end{IEEEproof}

From Lemma~\ref{lem:ave_TRU_dynTDD} and Theorem~\ref{thm:ave_TRU_statTDD},
we can further quantify the additional total TRU achieved by dynamic
TDD as
\begin{equation}
\kappa^{{\rm {ADD}}}=\frac{1}{T}\sum_{\tilde{k}=1}^{+\infty}\left[\left(p^{{\rm {U}}}\right)^{\tilde{k}}N_{0}^{{\rm {D}}}+\left(p^{{\rm {D}}}\right)^{\tilde{k}}N_{0}^{{\rm {U}}}\right]f_{\tilde{K}}\left(\tilde{k}\right),\label{eq:add_TRU_by_dynTDD}
\end{equation}
where $\kappa^{{\rm {ADD}}}$ measures the difference of $\kappa$
in (\ref{eq:Pr_DL_UL_Tx_dynTDD}) and that in (\ref{eq:Pr_DL_UL_Tx_statTDD}).%

In addition, we present the performance limit of $\kappa^{{\rm {ADD}}}$
in Lemma~\ref{lem:add_TRU_limit}.
\begin{lem}
\label{lem:add_TRU_limit}When $\lambda\rightarrow+\infty$, the limit
of $\kappa^{{\rm {ADD}}}$ is given by
\begin{equation}
\underset{\lambda\rightarrow+\infty}{\lim}\kappa^{{\rm {ADD}}}=\frac{p^{{\rm {D}}}N_{0}^{{\rm {U}}}}{T}+\frac{p^{{\rm {U}}}N_{0}^{{\rm {D}}}}{T}.\label{eq:add_TRU_by_dynTDD_limit}
\end{equation}
\end{lem}
\begin{IEEEproof}
From (\ref{eq:truncNB_PMF}), we have $\underset{\lambda\rightarrow+\infty}{\lim}\Pr\left[\tilde{K}=1\right]=1$.
Hence, using Lemma~\ref{lem:ave_TRU_dynTDD}, we can draw the following
conclusion for dynamic TDD:
\begin{equation}
\begin{cases}
\hspace{-0.2cm}\begin{array}{l}
\underset{\lambda\rightarrow+\infty}{\lim}\kappa^{{\rm {D}}}=1-f_{N^{{\rm {D}}}}\left(0\right)=1-p^{{\rm {U}}}=p^{{\rm {D}}}\\
\underset{\lambda\rightarrow+\infty}{\lim}\kappa^{{\rm {U}}}=1-f_{N^{{\rm {U}}}}\left(0\right)=1-p^{{\rm {D}}}=p^{{\rm {U}}}
\end{array}\end{cases}\hspace{-0.4cm}\hspace{-0.2cm}.\label{eq:Pr_DL_UL_Tx_dynTDD_limit}
\end{equation}
Based on $\underset{\lambda\rightarrow+\infty}{\lim}\Pr\left[\tilde{K}=1\right]=1$
and Theorem~\ref{thm:ave_TRU_statTDD}, we can obtain the following
conclusion for static TDD:
\begin{equation}
\begin{cases}
\hspace{-0.2cm}\begin{array}{l}
\underset{\lambda\rightarrow+\infty}{\lim}\kappa^{{\rm {D}}}=\left(1-\left(1-p^{{\rm {D}}}\right)\right)\frac{N_{0}^{{\rm {D}}}}{T}=\frac{p^{{\rm {D}}}N_{0}^{{\rm {D}}}}{T}\\
\underset{\lambda\rightarrow+\infty}{\lim}\kappa^{{\rm {U}}}=\left(1-\left(1-p^{{\rm {U}}}\right)\right)\frac{N_{0}^{{\rm {U}}}}{T}=\frac{p^{{\rm {U}}}N_{0}^{{\rm {U}}}}{T}
\end{array}\end{cases}\hspace{-0.4cm}.\label{eq:Pr_DL_UL_Tx_statTDD_limit}
\end{equation}
Our proof is completed by comparing (\ref{eq:Pr_DL_UL_Tx_dynTDD_limit})
with (\ref{eq:Pr_DL_UL_Tx_statTDD_limit}).
\end{IEEEproof}

Note that in (\ref{eq:add_TRU_by_dynTDD_limit}) of Lemma~\ref{lem:add_TRU_limit},
the first and the second terms are contributed from the DL and the
UL, respectively.

\section{Simulation and Discussion\label{sec:Simulation-and-Discussion}}

In this section, we present numerical results to validate the accuracy
of our analysis. In our simulations, the UE density is set to $\rho=300\thinspace\textrm{UEs/km}^{2}$,
which leads to $q=4.05$ in (\ref{eq:cov_area_size_PDF_Gamma})~\cite{Ding2016TWC_IMC}.%
\textbf{ }In addition, we assume that $p^{{\rm {D}}}=\frac{2}{3}$
and $T=10$. Thus, for static TDD, we have $N_{0}^{{\rm {D}}}=7$
and $N_{0}^{{\rm {U}}}=3$ in (\ref{eq:Pr_DL_UL_Tx_statTDD}), which
achieves the best match with $p^{{\rm {D}}}$ and $p^{{\rm {U}}}$,
according to (\ref{eq:DL_subf_selection_criterion}).%

\subsection{Validation of the Results on the Subframe Dependent TRU\label{subsec:subf_specific_TRU}}

\begin{figure}
\noindent \begin{centering}
\includegraphics[width=8cm]{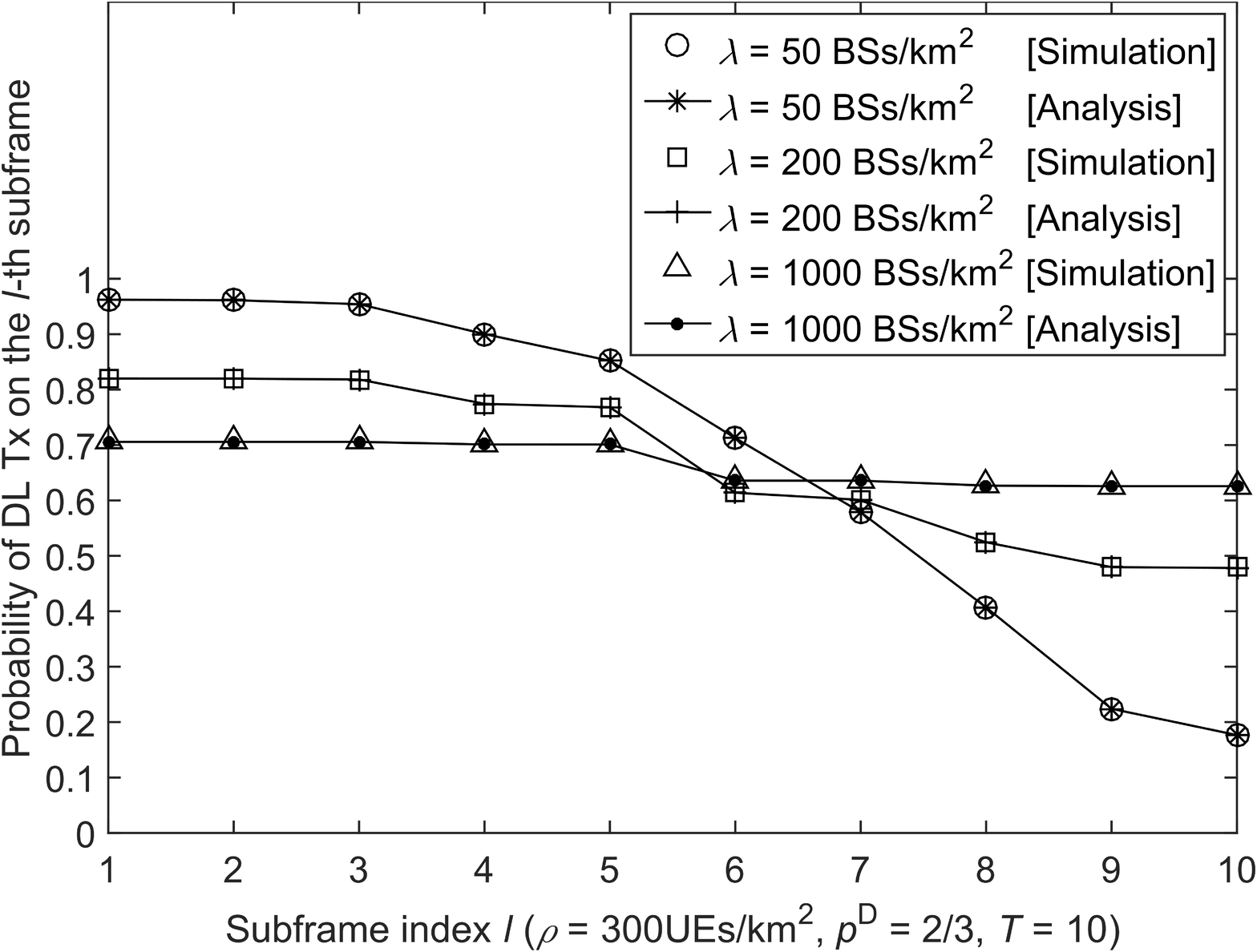}\renewcommand{\figurename}{Fig.}\caption{\label{fig:Pr_DL_Tx_on_l_subf_rho300_pD_0p66_TR2}The subframe dependent
DL TRU $q_{l}^{{\rm {D}}}$.}
\par\end{centering}
\vspace{-0.4cm}
\end{figure}

In Fig.~\ref{fig:Pr_DL_Tx_on_l_subf_rho300_pD_0p66_TR2}, we plot
the analytical and simulation results of $q_{l}^{{\rm {D}}}$.
{} For brevity, we omit showing the results of $q_{l}^{{\rm {U}}}$,
due to its duality to $q_{l}^{{\rm {D}}}$ as shown in (\ref{eq:condPr_subf_l_is_UL}).

From this figure, we can see that:
\begin{itemize}
\item The analytical results of $q_{l}^{{\rm {D}}}$ match well with the
simulation results. More specifically, the maximum difference between
the simulation results and the analytical results is shown to be less
than 0.3$\thinspace$\%. Note that the curves are not smooth in Fig.~\ref{fig:Pr_DL_Tx_on_l_subf_rho300_pD_0p66_TR2}.
This is because of the non-continuous mapping in (\ref{eq:DL_subf_selection_criterion}),
not due to an insufficient number of simulation experiments.
\item $q_{l}^{{\rm {D}}}$ monotonically decreases with $l$ because of
the considered DL-before-UL TDD configuration structure shown in Fig.~\ref{fig:TDD_configs}.
In other words, DL transmissions, if scheduled by BSs, are likely
to be conducted in earlier subframes.
\item When $\lambda$ is relatively small compared with $\rho$, e.g., $\lambda=50\thinspace\textrm{BSs/km}^{2}$,
$q_{1}^{{\rm {D}}}$ is almost one and $q_{10}^{{\rm {D}}}$ is close
to zero. This is because:
\begin{itemize}
\item the UE number in an active BS tends to be relatively large, e.g.,
the typical value of $\tilde{k}$ is around $4\thinspace\textrm{UEs/BS}$
when $\lambda=50\thinspace\textrm{BSs/km}^{2}$, as exhibited in Fig.~\ref{fig:PMF_UE_num_per_actBS_rho600_various_lambda};
\item as a result, the DL data request number is non-zero for most cases,
e.g., around $m^{{\rm {D}}}=3\thinspace$DL data requests when $p^{{\rm {D}}}=\frac{2}{3}$
and $\tilde{k}=4$, as shown in Fig.~\ref{fig:PMF_DL_dataReq_num_pD_0p66_various_k_tilde};
\item thus, the DL subframe number is also relatively large to accommodate
those DL data requests, e.g., the typical value of $n^{{\rm {D}}}$
is $\left\{ 3,5,8\right\} $ DL subframes when $p^{{\rm {D}}}=\frac{2}{3}$,
$\tilde{k}=4$ and $T=10$; as displayed in Fig.~\ref{fig:PMF_DL_subf_num_pD_0p66_various_k_tilde};
\item due to the DL-before-UL LTE TDD configuration structure shown in Fig.~\ref{fig:TDD_configs},
the first subframe has a very high probability to be a DL one, i.e.,
$q_{1}^{{\rm {D}}}$ is almost one, and the last subframe is the most
likely not to be a DL one, i.e., $q_{10}^{{\rm {D}}}$ is close to
zero.
\end{itemize}
\item When $\lambda$ is relatively large compared with $\rho$, e.g., $\lambda=1000\thinspace\textrm{BSs/km}^{2}$,
$q_{l}^{{\rm {D}}}$ is nearly a constant for all values of $l$.
Such constant roughly equals to the DL data request probability $p^{{\rm {D}}}=\frac{2}{3}$.
This is because:
\begin{itemize}
\item more than 80$\thinspace$\% of the active BSs will serve $\tilde{k}=1$
UE when $\lambda=1000\thinspace\textrm{BSs/km}^{2}$, as exhibited
in Fig.~\ref{fig:PMF_UE_num_per_actBS_rho600_various_lambda};
\item as a result, when $\tilde{k}=1$, both the DL data request number
and the DL subframe number show high fluctuation, as illustrated in
Figs.~\ref{fig:PMF_DL_dataReq_num_pD_0p66_various_k_tilde} and~\ref{fig:PMF_DL_subf_num_pD_0p66_various_k_tilde}
due to the strong traffic dynamics;
\item and hence, in most cases, all the subframes are used as either DL
ones or UL ones, the probability of which depends on the DL or UL
data request probability. Since $p^{{\rm {D}}}=\frac{2}{3}$, we have
that $q_{l}^{{\rm {D}}}\approx\frac{2}{3}$.
\end{itemize}
\item It is straightforward to see that the case of $\lambda=200\thinspace\textrm{BSs/km}^{2}$
plays in the middle of the above two cases%
.
\end{itemize}

\subsection{Validation of the Results on the Average TRU\label{subsec:average_TRU}}

\begin{figure}
\noindent \begin{centering}
\includegraphics[width=8cm]{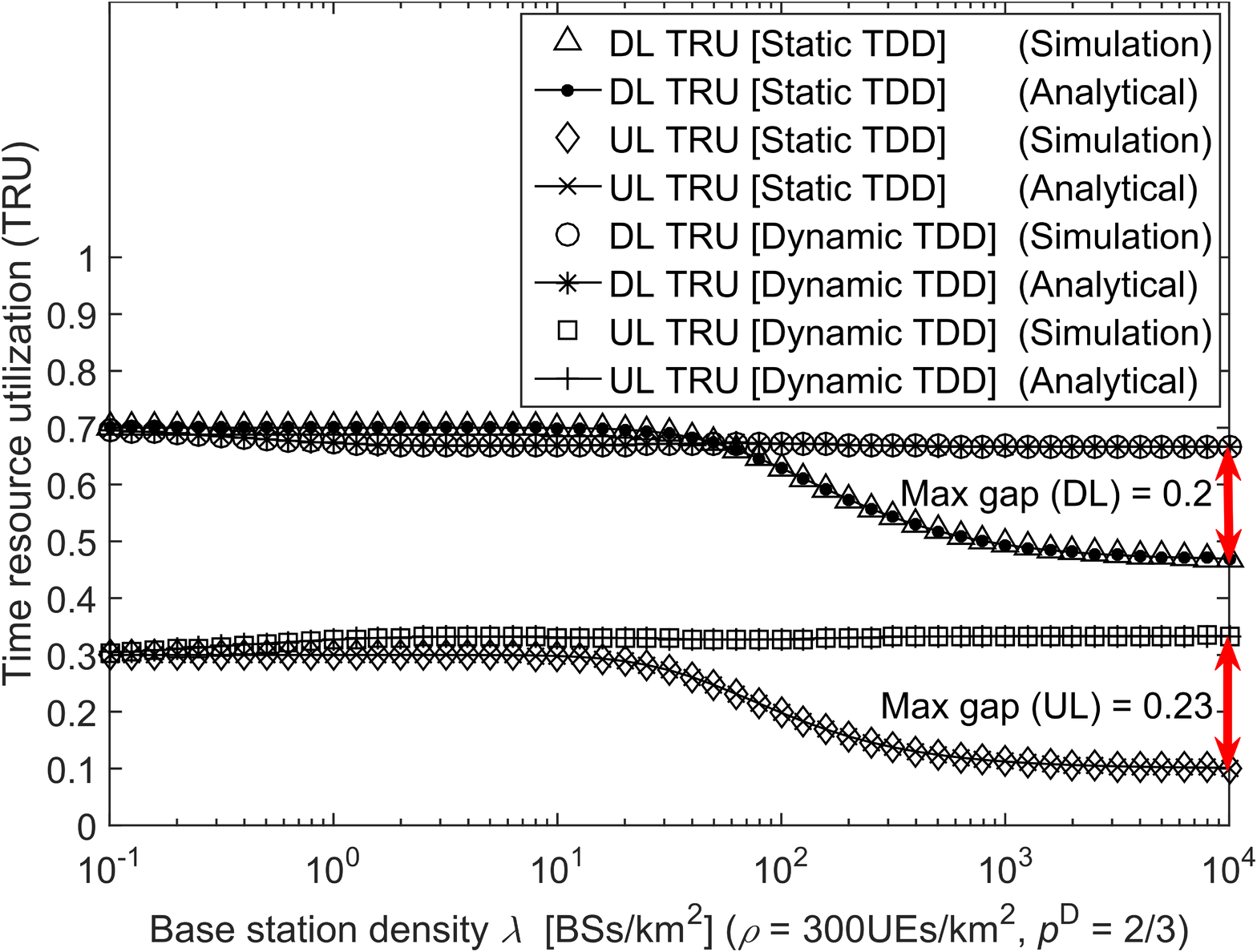}\renewcommand{\figurename}{Fig.}\caption{\label{fig:TRU_pD_0p66}The average DL/UL TRU $\kappa^{{\rm {D}}}$
and $\kappa^{{\rm {U}}}$.}
\par\end{centering}
\vspace{-0.4cm}
\end{figure}

In Fig.~\ref{fig:TRU_pD_0p66}, we display the analytical and simulation
results of $\kappa^{{\rm {D}}}$ and $\kappa^{{\rm {U}}}$.
{} From this figure, we can see that:
\begin{itemize}
\item The analytical results of $\kappa^{{\rm {D}}}$ and $\kappa^{{\rm {U}}}$
match well with the simulation results for both static TDD and dynamic
TDD.
\item The sum of $\kappa^{{\rm {D}}}$ and $\kappa^{{\rm {U}}}$, i.e.,
$\kappa$, for dynamic TDD is shown to always equal to one, which
verifies Lemma~\ref{lem:ave_TRU_dynTDD}, and shows the main advantage
of dynamic TDD.
\item The $\kappa$ for static TDD become smaller than one in dense SCNs,
e.g., $\lambda>10\thinspace\textrm{BSs/km}^{2}$, which verifies Theorem~\ref{thm:ave_TRU_statTDD},
and shows the main drawback of static TDD.%
\item According to Lemma~\ref{lem:add_TRU_limit}, the additional total
TRU achievable by dynamic TDD, $\underset{\lambda\rightarrow+\infty}{\lim}\kappa^{{\rm {ADD}}}$,
should be composed of two parts, i.e., one from the DL ($\frac{p^{{\rm {D}}}N_{0}^{{\rm {U}}}}{T}$)
and another one from the UL ($\frac{p^{{\rm {U}}}N_{0}^{{\rm {D}}}}{T}$),
which in this case are 0.2 and 0.23, respectively. These values can
be validated in Fig.~\ref{fig:TRU_pD_0p66}. Thus, the average total
gain of dynamic TDD over static TDD in terms of $\underset{\lambda\rightarrow+\infty}{\lim}\kappa^{{\rm {ADD}}}$
is 0.43 (+$75.4\,\%$).%
\end{itemize}

\section{Conclusion\label{sec:Conclusion}}

For the first time, we investigated the MAC layer TRU for synchronous
dynamic TDD with the network densification, which quantifies the performance
of dynamic TDD in terms of its MAC layer resource usage. We showed
that the DL/UL TRU varies across TDD subframes, and that dynamic TDD
can achieve an increasingly higher average total TRU than static TDD
with the network densification of up to $75.4\,\%$. As future work,
we will combined our results on the MAC layer TRU with the PHY layer
SINR results to derive the total area spectral efficiency for synchronous
dynamic TDD.%

\bibliographystyle{IEEEtran}
\bibliography{Ming_library}

\end{document}